\documentclass[%
 aip,
 amsmath,amssymb,
 reprint,%
]{revtex4-1}

\usepackage{graphicx}
\usepackage{dcolumn}
\usepackage{bm}
\usepackage{textcomp}
\usepackage{color}

\usepackage[utf8]{inputenc}
\usepackage[T1]{fontenc}
\usepackage{mathptmx}

\begin{document}

\preprint{AAPM/123-QED}

\title{Fast control of dissipation in a superconducting resonator}

\author{V.~A.~Sevriuk}
 \affiliation{QCD Labs, QTF Centre of Excellence, Department of Applied Physics, Aalto University, PO Box 13500, FI-00076 Aalto, Finland}
\author{K.~Y.~Tan}%
\affiliation{QCD Labs, QTF Centre of Excellence, Department of Applied Physics, Aalto University, PO Box 13500, FI-00076 Aalto, Finland
}%
\author{E.~Hyypp\"a}
 \affiliation{QCD Labs, QTF Centre of Excellence, Department of Applied Physics, Aalto University, PO Box 13500, FI-00076 Aalto, Finland}
  \author{M.~Silveri}
 \affiliation{QCD Labs, QTF Centre of Excellence, Department of Applied Physics, Aalto University, PO Box 13500, FI-00076 Aalto, Finland}
 \affiliation{Research unit of nano and molecular systems, University of Oulu, PO Box 3000, FI-90014 Oulu, Finland}
 
 \author{M.~Partanen}
 \affiliation{QCD Labs, QTF Centre of Excellence, Department of Applied Physics, Aalto University, PO Box 13500, FI-00076 Aalto, Finland}
  \author{M.~Jenei}
 \affiliation{QCD Labs, QTF Centre of Excellence, Department of Applied Physics, Aalto University, PO Box 13500, FI-00076 Aalto, Finland}
   \author{S.~Masuda}
 \affiliation{QCD Labs, QTF Centre of Excellence, Department of Applied Physics, Aalto University, PO Box 13500, FI-00076 Aalto, Finland}
   \author{J.~Goetz}
 \affiliation{QCD Labs, QTF Centre of Excellence, Department of Applied Physics, Aalto University, PO Box 13500, FI-00076 Aalto, Finland}
   \author{V.~Vesterinen}
 \affiliation{VTT Technical Research Centre of Finland Ltd, P.O. Box 1000, FI-02044, VTT, Finland}
    \author{L.~Gr\"onberg}
 \affiliation{VTT Technical Research Centre of Finland Ltd, P.O. Box 1000, FI-02044, VTT, Finland}
\author{M.~M\"ott\"onen}
 \affiliation{QCD Labs, QTF Centre of Excellence, Department of Applied Physics, Aalto University, PO Box 13500, FI-00076 Aalto, Finland}
 \affiliation{VTT Technical Research Centre of Finland Ltd, P.O. Box 1000, FI-02044, VTT, Finland}

\date{\today}

\begin{abstract}
	We report on fast tunability of an electromagnetic environment coupled to a superconducting coplanar waveguide resonator. Namely, we utilize a recently-developed quantum-circuit refrigerator (QCR) to experimentally demonstrate a dynamic tunability in the total damping rate of the resonator up to almost two orders of magnitude. Based on the theory it corresponds to a change in the internal damping rate by nearly four orders of magnitude. The control of the QCR is fully electrical, with the shortest implemented operation times in the range of 10~ns. This experiment constitutes a fast active reset of a superconducting quantum circuit. In the future, a similar scheme can potentially be used to initialize superconducting quantum bits.
\end{abstract}

\maketitle

Tunable dissipative environments for circuit quantum electrodynamics (cQED) are pursued intensively in experiments due to the unique opportunities to study non-Hermitian physics~\cite{Leggett1987,Rastelli2018}, such as phase transitions related to parity-time symmetry~\cite{pttheory2018}, decoherence and quantum noise~\cite{Clerk10}. Interesting effects can be observed in experiments on exceptional points~\cite{Partanen18,sweeney2019,Heiss2012}, which also gives possibilities to use such systems as models in nonlinear photonics, for example, for metamaterials~\cite{Cui13} and for photonic crystals~\cite{Kim2016}.

From the practical point of view, tunable environments are utilized to protect and process quantum information~\cite{Mirrahimi_2014,kraus2008,Verstraete2009} and to implement qubit reset~\cite{Jones2013,Wendin2017}. The latter application calls for fast tunability of the environment due to the aim to increase the rate of the operations on a quantum computer. Recent advances in the field of cQED for quantum information processing~\cite{Wendin2017,Divincenzo00,Devoret1169,reagor2016} render this topic highly interesting. 

\begin{figure}
  \centering
  \includegraphics[width=0.9\linewidth]{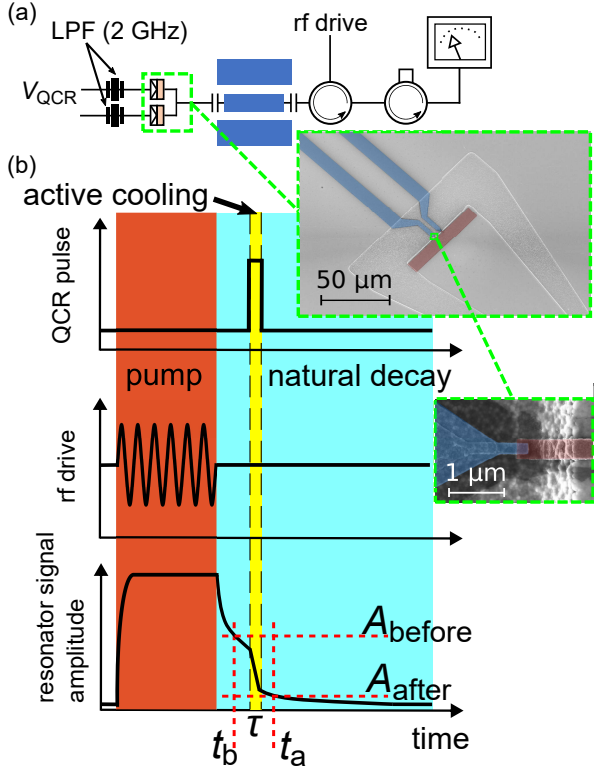}
  \caption{\label{fig:scheme}(a) Simplified measurement setup. The sample consists of a coplanar waveguide resonator (blue) capacitively coupled to a QCR (on the left) and a transmission line (on the right). The QCR is represented by two NIS junctions, SEM images of which are shown in the insets, where blue color denotes superconductor and red stands for normal-metal island. The superconducting leads of the QCR are connected through two on-chip low-pass filters to transmission lines. (b) Measurement protocol. Red color indicates the time interval for driving the resonator with an rf tone. Then we switch the drive off and observe the decay of the resonator signal (light blue region). At a chosen point of time (yellow region) we send a voltage pulse to the QCR with width $\tau$  to induce active cooling. To extract the damping rate of the QCR we choose two points in time, before ($t_\mathrm{b}$) and after ($t_\mathrm{a}$) the QCR pulse and measure the corresponding amplitudes of the resonator signal.}
\end{figure}

There are different ways for resetting superconducting qubits. Firstly, one may tune the qubit frequency to reduce its life time~\cite{resetReed2010}. The disadvantages of this method include the broad frequency band reserved by the qubit and the required fast frequency sweep which may lead to an increased amount of initialization error~\cite{tuorila2019}. Conventionally, it is beneficial to maintain the qubits at the optimal parameter points during all operations. Secondly, it is possible to use microwave pulses to actively drive the qubit to the ground state~\cite{resetGeerlings2013,pulseresetEgger2018,resetMagnard2018,Valenzuela2006}. Such methods are popular because no additional components or new control steps are needed. However, to achieve high fidelity one usually needs to increase the reset time to the microsecond range. Thirdly, one can engineer a tunable environment for the qubits~\cite{Jones2013,basilewitsch2019,Tuorila2017,Wong_2019}. This approach demand changes in the chip design, but may lead to high fidelity for a fast reset without compromises on the other properties.

Here, we focus on a single-parameter-controlled tunable environment implemented by a quantum-circuit refrigerator (QCR)~\cite{Partanen18, Tan2017, Silveri17prb, Masuda2018, hyyppa2019, Silveri2019}. The refrigerator is based on photon-assisted electron tunneling through two identical normal-metal--insulator--superconductor junctions (NIS). It has been used to cool down a photon mode of the resonator~\cite{Tan2017}, and to observe a Lamb shift in a cQED system~\cite{Silveri2019}. Furthermore, QCR can be used as a cryogenic photon source~\cite{Masuda2018}, which makes it applicable for a calibration of cryogenic amplification chains~\cite{hyyppa2019}. Details of the QCR operation and theory are described in Ref.~\onlinecite{Silveri17prb}. 

The QCR can potentially be incorporated in a quantum processor to significantly decrease the reset time. However, all previous related research has been based on the dc and low-frequency voltage control of the QCR-induced dissipation rates. In this paper, in contrast, we demonstrate nanosecond operation of the QCR, which serves to expand future application areas of the this device.

As shown in Fig.~\ref{fig:scheme}(a), the measured sample consists of a coplanar waveguide resonator, coupled to an rf transmission line at one end and to a normal-metal island of the QCR at the other end. The superconducting leads of the QCR are connected to two transmission lines with on-chip low-pass filters. The low-pass filters were introduced in this batch of samples for the first time in the context of the QCR and used to reduce the high-frequency noise. The filters are realized as lumped-element RC elements made of palladium fins capacitively shunted to ground through an aluminum oxide layer. All the measurements are carried out in a dilution refrigerator with the base temperature of 10 mK. 

The fabrication of the sample is fully described in our previous publication~\cite{Masuda2018}. In brief, the sample is fabricated on a 500-\textmu{}m-thick Si wafer. The resonator is defined in a 200-nm-thick sputtered niobium layer, and subsequently covered by a 50-nm-thick layer of $\mathrm{Al_2O_3}$. The NIS junctions are defined with electron beam lithography followed by two-angle evaporation (20~nm Al, 20~nm Cu).

We provide a summary of the relevant sample parameters in Tab.~\ref{t1}. The parameters are estimated from the fabrication process and previous measurements from similar samples~\cite{Silveri2019}. The SEM in Fig.~\ref{fig:scheme} are obtained on a sample from the same fabrication batch as the sample, which is used in this work.

For the QCR characterization, we use homodyne detection to measure the amplitude of the resonator signal voltage from the connected transmission line. A schematic example of the measurement result together with the basic sequence of the applied pulses are shown in Fig.~\ref{fig:scheme}(b). First, we send an rf drive to the resonator to populate the fundamental mode. The power of the resonator rf drive is around --110 dBm at the sample input. Then we switch the drive off and observe the decay of the resonator signal. At this stage there are three main sources of decay: coupling to the transmission line, to the QCR in the off state, and to unknown excess sources. The value of the damping rate related to the unknown sources $\gamma_{\mathrm{x}}$ is estimated from the previous measurements~\cite{Silveri2019} to be roughly 10\% of the damping rate to the transmission line $\gamma_{\mathrm{tr}}$. The QCR-off damping rate $\gamma_{\mathrm{QCR}}^{\mathrm{off}}$ is negligible, in line with previous studies~\cite{Silveri2019}. At a chosen instant of time, we send a voltage pulse to the QCR, which changes the rate of the resonator signal decay. The QCR voltage pulse is formed by an arbitrary waveform generator.

\renewcommand{\arraystretch}{1.5}
\begin{table}
  
  \caption{\label{t1}Parameters of the measured sample.}
  \begin{ruledtabular}

 \begin{tabular}{c c c}
  
    $R_\mathrm{T}$ & $14$ $\mathrm{k\Omega}$ & NIS tunneling resistance  \\ 
       \hline 
    $T_\mathrm{N}$ & 0.17 K & electron temperature of the normal-metal island \\ 
       \hline        
    $\gamma_\mathrm{D}$ & $4\times10^{-4}$ & Dynes parameter \\ 
       \hline 
    $Z_\mathrm{r}$ & 35  $\mathrm{\Omega}$ & resonator impedance \\ 
       \hline 
    $C_\mathrm{c}$ & $840 $ fF & capacitance between QCR and resonator  \\ 
       \hline 
    $C_\mathrm{m}$ & $5 $ fF & NIS junction capacitance \\ 
       \hline 
    $f_\mathrm{0}$ & 8.683 GHz & resonator mode frequency \\ 
       \hline 
    $\mathrm{\Delta}$ & 215 $\mathrm{\mu{}}$eV & energy gap parameter of the Al leads \\

  \end{tabular}
  \end{ruledtabular}

\end{table}

To estimate the damping rate arising from the QCR during the voltage pulse (on state), we use a procedure, explained below. It is based on varying both the amplitude and the width of the QCR pulse in an effort to increase the precision of the measurements.

First, we choose two points in time, one before and one after the QCR pulse, see Fig.~\ref{fig:scheme}(b). These points are fixed for all subsequent measurements. We express the amplitude of the resonator signal after the QCR pulse as
\begin{eqnarray}
A_{\mathrm{after}} = A_{\mathrm{before}}   \mathrm{exp}\{-\frac{1}{2}[\gamma_{\mathrm{QCR}}  (\tau-\Delta t_{\mathrm{rise}}-\Delta t_{\mathrm{fall}})\nonumber \\+ (\gamma_{\mathrm{tr}}+\gamma_{\mathrm{x}})(t_\mathrm{a}-t_\mathrm{b}) +\gamma_{\mathrm{QCR}}^{\mathrm{rise/fall}}  (\Delta t_{\mathrm{rise}}+\Delta t_{\mathrm{fall}})\nonumber \\+ \gamma_{\mathrm{QCR}}^{\mathrm{off}}(t_\mathrm{a}-t_\mathrm{b}-\tau)]\},
\end{eqnarray}

\noindent
where $A_{\mathrm{after}}$ and $A_{\mathrm{before}}$ are the amplitudes of the resonator signal voltage after ($t_\mathrm{a}$) and before ($t_\mathrm{b}$) the QCR pulse, $\tau$ is the width of the pulse, $\gamma_{\mathrm{QCR}}$ is the damping rate arising from the QCR in the on state, excluding the times of the rise $\Delta t_{\mathrm{rise}}$ and fall $\Delta t_{\mathrm{fall}}$ of the QCR voltage, and $\gamma_{\mathrm{QCR}}^{\mathrm{rise/fall}}$ is the effective  damping rate during the pulse rise and fall. In this work, we aim to extract $\gamma_{\mathrm{QCR}}$ which is is the most relevant parameter for fast circuit reset and can be connected to the theory, previous measurements, and future applications of the QCR. 

\begin{figure}
  \centering
  \includegraphics[width=0.98\linewidth]{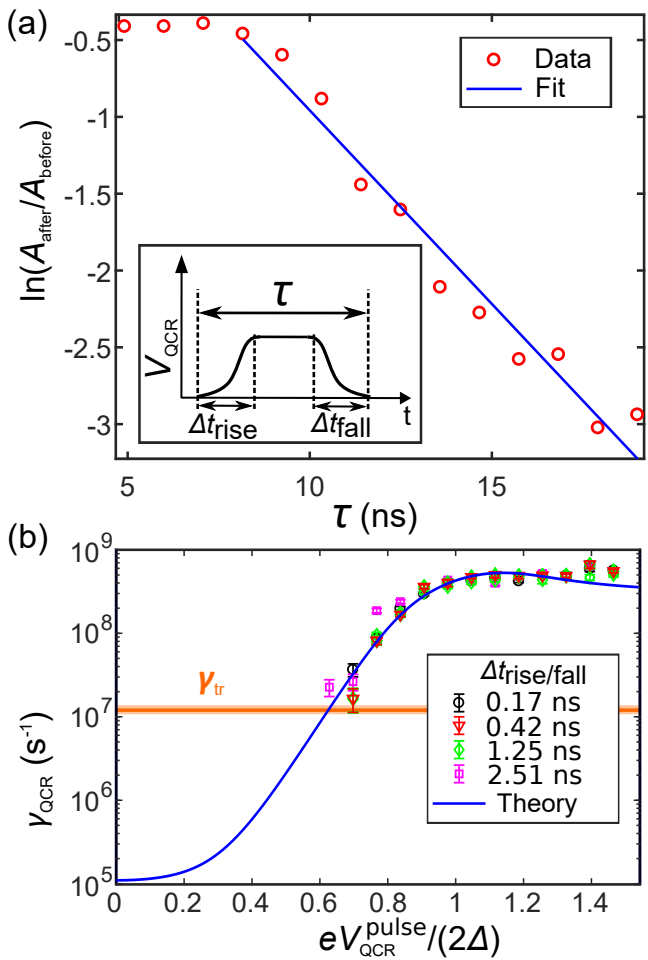}
  \caption{\label{fig:res}(a) Logarithm of the ratio of the measured output resonator signal amplitudes before and after the QCR pulse as a function of the pulse width (circles) and a linear fit (solid line) which is used to extract the damping rate of the QCR ($\gamma_{\mathrm{QCR}}$). The inset shows the shape of the QCR pulse such that $\Delta t_{\mathrm{rise/fall}}$ are included in $\tau$ and denote the width of the sine-squared functions which are used for the rise and fall of the QCR voltage. In this example $V^{\mathrm{pulse}}_{\mathrm{QCR}}=345$ $\mathrm{\mu}{}V$ ($0.8 \times 2\Delta$) and $\Delta t_\mathrm{rise/fall}=1.25$~ns at the output of the arbitrary waveform generator. (b) Damping rate of the QCR $\gamma_{\mathrm{QCR}}$ as a function of the amplitude of the voltage pulse applied to the QCR. We normalize the amplitude by the energy gap of the superconducting leads, which is $2\Delta=2\times215$~$\mathrm{\mu}$eV. The difference between the data sets is the different times for the rise and fall of the QCR voltage as indicated. Orange line with shaded region denotes the damping rate of the transmission line $\gamma_{\mathrm{tr}}$ together with its 1$\sigma$ uncertainty. The error bars of the experimental data points denote the 1$\sigma$ uncertainties arising from the corresponding linear fit.}
\end{figure}

Next, we vary $\tau$ such that it is still smaller than $t_\mathrm{a}-t_\mathrm{b}$. Only two components in the exponential function are changing with this variation, namely, those related to $\gamma_{\mathrm{QCR}}$ and $\gamma_{\mathrm{QCR}}^{\mathrm{off}}$. The latter, as mentioned above, is negligibly small. Thus ln($A_{\mathrm{after}}$/$A_{\mathrm{before}}$) is a linearly decreasing function of $\tau$ with a slope $\gamma_{\mathrm{QCR}}/2$, see Fig.~\ref{fig:res}(a). In this figure, we observe a flat region at short pulses, which is related to the rise and fall of the QCR voltage. The flat region extends to approximately 8~ns, which is significantly longer than $\Delta t_{\mathrm{rise}}+\Delta t_{\mathrm{fall}}=2.50$~ns. This effect can be explained by the distortions of the voltage pulse due to the imperfection of the QCR control line. The width of the flat region is varying depending on the height of the voltage pulses and the sum of the rise and fall times (data not shown). Importantly, a clear linearly decreasing  part of the graph yields the damping rate of the QCR during the pulse.

Figure~\ref{fig:res}(b) shows the extracted damping rate of the QCR, $\gamma_{\mathrm{QCR}}$, as a function of the height of the applied QCR voltage pulse $V^{\mathrm{pulse}}_{\mathrm{QCR}}$. There are four different data sets, corresponding to the different $\Delta t_{\mathrm{rise/fall}}$, which are varied in an effort to reduce the switching time of the QCR. The shortest measured sum of the rise and fall times is 6~ns, which was obtained with nominal $\Delta t_{\mathrm{rise}}+\Delta t_{\mathrm{fall}}=0.34$~ns and $V^{\mathrm{pulse}}_{\mathrm{QCR}}=345$ $\mathrm{\mu}{}V$ ($0.8 \times 2\Delta$). The results in Fig.~\ref{fig:res}(b) follow well the theoretical prediction. There are no experimental data points lower than $1.6\times{}10^7$ $\mathrm{s}^{-1}$. The difficulty to measure in this range is explained by the fact that the damping rate of the transmission line $\gamma_{\mathrm{tr}}$ is around this value, and hence dominates the decay at the voltages lower than 0.6$\times 2 \Delta$. This decreases the signal-to-noise ratio in the extraction procedure for $\gamma_{\mathrm{QCR}}$. The damping rate of the transmission line $\gamma_{\mathrm{tr}}$ is calculated from the measured decay of the resonator signal prior to the QCR voltage pulse.
Short list of the specific characteristic values of the extracted damping rates are shown in Tab.~\ref{t2}. 

\renewcommand{\arraystretch}{1.5}
\begin{table}
  
  \caption{\label{t2}Principle characteristic values of the obtained damping rates, see Fig.~\ref{fig:res}(b), including maximal $\gamma^{\mathrm{max}}_{\mathrm{QCR}}$ and minimal $\gamma^{\mathrm{min}}_{\mathrm{QCR}}$ measured damping rate of the QCR, theoretical value of the QCR damping rate at zero bias voltage $\gamma^{\mathrm{off}}_{\mathrm{QCR}}$ and measured damping rate of the transmission line $\gamma_{\mathrm{tr}}$.}
  \begin{ruledtabular}

 \begin{tabular}{c c c}
  
   $\gamma^{\mathrm{max}}_{\mathrm{QCR}}$ & $6.7\pm 0.7\times 10^{8}$ $\mathrm{s^{-1}}$ & measured  \\ 
       \hline 
   $\gamma^{\mathrm{min}}_{\mathrm{QCR}}$ & $1.6\pm 0.5\times 10^{7}$ $\mathrm{s^{-1}}$ & measured \\ 
       \hline     
   $\gamma^{\mathrm{off}}_{\mathrm{QCR}}$ & $1.1\times 10^{5}$ $\mathrm{s^{-1}}$ & theory  \\ 
       \hline         
   $\gamma_{\mathrm{tr}}$ & $1.2\pm 0.1\times 10^{7}$ $\mathrm{s^{-1}}$ & measured  \\

  \end{tabular}
  \end{ruledtabular}

\end{table}

In our experiment the total damping rate of the resonator can be tuned up to factor of $56\pm8$, which is calculated from the ratio $\gamma^{\mathrm{max}}_{\mathrm{QCR}}/\gamma_{\mathrm{tr}}$. Based on the theory~\cite{Silveri17prb}, however, we estimate that the damping rate of the QCR changes by almost four orders of magnitude. Thus in our experiment the efficiency of the QCR is limited by the strong coupling of the resonator to the transmission line. However, our observations support the abovementioned scenario of resetting superconducting qubits. The measurement results on $\gamma_{\mathrm{QCR}}$ in Fig.~2(b) suggest that the QCR is giving very small changes of the damping rate in the off state and an optimized QCR voltage pulse can reduce the number of the resonator photons to less than 1\% of the original number in a few tens of nanoseconds.

In conclusion, we experimentally demonstrated that a QCR can be turned on and off in a nanosecond time scale. This renders it as a  potentially useful device in resetting a multitude of quantum electric circuits. In the future, we aim to couple the QCR with transmon qubits which allows a more accurate measurement of the on/off ratio and the resulting temperature of the refrigerated system. 

\section*{ACKNOWLEDGMENTS}
This project has received funding from the European Union’s Horizon 2020 Research and Innovation Programme under the Marie Sk\l{}odowska-Curie Grant No.~795159 and under the European Research
Council Consolidator Grant No. 681311 (QUESS), from
the Academy of Finland Centre of Excellence in Quantum Technology Grant No. 312300, No. 316619. and No. 305237, from
JST ERATO Grant No. JPMJER1601, from JSPS KAKENHI Grant No. 18K03486, from the EU Flagship project QMiCS, from the Emil Aaltonen Foundation, from the Alfred Kordelin Foundation,  and from the Vilho, Yrj\"o and
Kalle V\"ais\"al\"a Foundation. We acknowledge the provision
of facilities and technical support by Aalto University at
OtaNano - Micronova Nanofabrication Centre.

\bibliography{PE_refs.bib}{}
\end{document}